# Using Natural Language Processing and Qualitative Analysis to Intervene in Gang Violence: A Collaboration Between a Social Work Researchers and Data Scientists

Desmond Upton Patton
Columbia University
New York, NY, USA
dp2787@columbia.edu

Kathleen McKeown
Columbia University
New York, NY, USA
Kathy@cs.columbia.edu

Owen Rambow
Columbia University
New York, NY, USA
Ocr2101@columbia.edu

Jamie Macbeth
Fairfield University
Fairfield, CT, USA
Jamie.Macbeth@gmail.com

**ABSTRACT**

The U.S. has the highest rate of firearm-related deaths when compared to other industrialized countries. Violence particularly affects low-income, urban neighborhoods in cities like Chicago, which saw a 40% increase in firearm violence from 2014 to 2015 to more than 3,000 shooting victims. While recent studies have found that urban, gang-involved individuals curate a unique and complex communication style within and between social media platforms, organizations focused on reducing gang violence are struggling to keep up with the growing complexity of social media platforms and the sheer volume of data they present.  In this paper, describe the Digital Urban Violence Analysis Approach (DUVVA), a collaborative qualitative analysis method used in a collaboration between data scientists and social work researchers to develop a suite of systems for decoding the high- stress language of urban,  gang-involved youth. Our approach leverages principles of grounded theory when analyzing approximately
800 tweets posted by Chicago gang members and participation of youth from Chicago neighborhoods to create a language resource for natural language processing (NLP) methods. In uncovering the unique language and communication style, we developed automated  tools with the potential to detect aggressive language on social media and aid individuals and groups in performing violence prevention and interruption.

## 1. INTRODUCTION

In this paper, we describe the Digital Urban Violence Analysis approach (DUVVA), a collaborative qualitative analysis method used to inform natural language processing techniques that predict clusters of aggressive language on Twitter that may escalate into fatal and non-fatal firearm violence. Recently, several studies (Patton et al 2015; Patton et al. 2016a; Patton et al. 2016b) examined the relationship between social media and gang violence. These study found that young people living in violent, urban neighborhood taunt each other, make threats and brag about violence on social media platforms in ways that may lead to firearm violence, a behavior known as *Internet banging or cyberbanging* (Patton, Eschmann, Butler, 2013). These studies used in-depth, qualitative methods that include coding each post by hand to code text from known gang-involved youth in Chicago. However, this method alone is inefficient when analyzing massive amounts of social media data that requires an in-depth understanding of community context and language.

We aim to develop a more efficient, automatic coding process using natural language processing (NLP) tools informed by in-depth qualitative insights of offline conditions and mechanisms that shape urban gang-related violence. The development of the NLP tool could be particularly helpful for violence prevention organizations that use social media as a part of their violence intervention methods.

One study of Chicago gang communication on Twitter found that urban, gang-involved individuals curate a unique, complex communication style within and between social media platforms that warrant careful interpretation. Many studies of Twitter are "big data" studies that employ quantitative computer-based analysis of tens of thousands or even millions of data points. However, there are many obstacles to collecting quantitative data of gang-related behavior on Twitter. Identifying hashtags or keywords is difficult given the diverse kinds of users and communication patterns and styles that vary by city, neighborhood and between gangs, crews, and cliques. Furthermore, the linguistic style and extensive use of emoji's and other images among gang-affiliated youth renders most quantitative tools such as scripts or emotion detection software inadequate. There is a need to understand more accurately how social media reflects the lived reality of marginalized young people who live in low-income, violent urban neighborhoods. To achieve this goal, we must first understand how language is used to communicate offline identities, networks and exposure to violence and trauma. With its' great popularity among youth, social media is an understudied youth environment where implicitly or explicitly, beliefs and attitudes are shared, norms transmitted, and behavior modeled.

## 2. CORPUS

We analyzed the Twitter communication of Gakirah Barnes and her most frequent communicators. Gakirah Barnes was a 17-year female gang member from Chicago who created the online Twitter account @TyquanAssassin in memory of her friend Tyquan Tyler who was allegedly killed by a rival gang in 2013. We chose to focus on Gakirah because she was active on Twitter posting over 27,000 tweets from December 2011 until her own death on April 11, 2014. She used her account to express the events of her daily life, which ranged from friendships and other relationships to gang violence, and grieving the death of her peers to gun violence. We used Radian6, a social media tracking service, to capture an overall sample of 10,000 tweets by, mentions of, and replies to @TyquanAssassin. We apply a close, textual read of 800 of the 10,0000 tweets during a two-week period. We begin our analysis with the day in which Gakirah's friend, Raason "Lil B" Shaw, was allegedly killed by the Chicago police (March 29th, 2014) to one week after Gakirah's death (Thursday, April 17th, 2014).

## 3. HUMAN ANNOTATION OF DATA

We created an interdisciplinary team of researchers which consisted of two Chicago youth, a social work professor, two masters-level social work students; three data scientist along with two undergraduate data science students (Ford, 2014). We describe DUVAA and apply it to building a natural language processing tool to predict clusters of aggression and loss among gang-involved youth in Chicago.

### 3.1. Qualitative Analysis

To begin our analysis, we hired two young men of color from a Chicago neighborhood with high rates of violence. These young men were asked to carefully translate 800 tweets from Gakirah Barnes during a two-week period following two homicides: the death of Gakirah's close friend and then her subsequent death two weeks later. The Chicago research assistants were provided an excel spreadsheet with Gakirah's twitter data which listed the author of the tweet, the content (excluding images) and the URL for the specific Twitter page linked to the author. The Chicago research assistants then provided their initial translations of the tweets to include them describe: their first reactions to reading the tweet, their understanding of the tone and emotionality of the tweet and then an explanation of language used. The research assistants also interpreted emoji's that were connected to text when they were able to access the URL for a specific tweet.

The social work team, comprised for two Master of Social Work students and a social work professor) used the translations from the Chicago research assistants as the initial training data to develop a social media corpus that provided an in-depth and accurate understanding of the culture, context and language embedded in the tweets. Using a random sample of 50 Twitter communications from Gakirah and users in her Twitter network, a codebook was developed to reflect 26 categories or themes found in the data The research team then used the codebook to code approximately 800 tweets. Some examples of codes included aggression, grief, threats and discussions of money and relationships

Next, informed by Chicago research assistant translations, the social work team took a deeper look at the Twitter data and examined posted labeled as aggression or threat and asked: "why was this communicated?" To do this, we used created the Digital Urban Violence Analysis Approach (DUVAA) a six step qualitative analysis approach to examining urban violence related content on social media. First, we identified the *Precipitating Events* that may trigger an aggressive or threatening conversation. A precipitating event may include a death/homicide, dispute over territory or a relationship that may have turned negative. We then look at the *Twitter Author* of the tweet. We review their Twitter handle within the context of their overall Twitter profile. We look for clues that suggest where they are from, and who their friends are, to better understand how they present themselves on Twitter. We then examine the *Content* of the tweet. For example, we looked for individuals that may have been referenced in a tweet as that might the authors' social network. We take note of names of gangs or affiliations with crews or cliques, and hashtags that may indicate location, gang affiliation, neighborhood, mood, etc.

The content offers additional *Clues* that describe offline characteristics that may explain how and or why a Twitter post can lead to violence. Clues include conversations, replies, and mentions that provide information about the author and who is in their social network. For example, we pay attention to grammatical constructions, specifically, capitalization, the use of punctuation. We look at abbreviations, and numerals as well. Next, we examined the *Tone* of the tweets. We identified jokes, irony, references to products (e.g., promoting videos), and rap lyrics. Our final step involved identifying *Trigger Events*. Here we identified points in the Twitter narrative where communication shifts from general or positive communication to aggressive of threatening communication. For example, we look for the introduction of a new offline event or experience (e.g., a friend is killed by a rival group, someone goes to jail, language shifts from pleasant to aggressive, etc.). Central to this process was acquiring additional context for Gakirah and users in her Twitter network. We reviewed the biographical blurb on her Twitter profile and reviewing images and videos that may highlight additional connections to gangs, the location of activities, and participation in leisurely activities like creating rap videos (Lewis et al. 2013). During this process, we acquired a deeper understanding of the context surrounding the variation in Twitter communication. For example, we learned that aggressive and threatening communication often preceded by posts that reflected loss or grief of a loved one due to gun violence.

We went back to the data set and re-coded the twitter data based on our new insights informed by DUVAA. Imperative to this process was the coding meetings where the social work team came together with the data scientists and Chicago research assistant to discuss how the data was coded. For example, the Chicago research assistants would clarify new terms, language or expressions uncovered in the dataset that derived from DUVAA coding. During the meetings the social work, the team described how they developed codes and identified emerging themes. Also, the data scientists asked questions about the qualitative coding process to understand better why some tweets were coded as aggression/threat or grief.

Based on the coding meetings, we then engaged in the second round of coding or selective coding which was used to examine further why a category existed. We produced the "test" data set during the second round of coding. We collapsed our 26 codes further based on an observable pattern in the data that suggested our codes fit into three broad categories: 1) aggression, 2) grief and 3) other based on frequency. The data set was coded by the two master of social work students; inter-annotator agreement on the test data set is K=0.62, which is moderate agreement.
The collapsed aggression code contained examples of insults, threats, bragging, hypervigilance and challenges with authority. The collapsed grief code includes examples of: distress, sadness, loneliness, and death. The "other" codes contained examples of general conversations between users, discussions about women, and tweets that represented happiness. The data scientist team then used the collapsed codes to predict clusters of aggressive and grief communication within Gakirah's Twitter communication.

### 3.2 Natural Language Processing

In this stage of the work, we used natural language processing (NLP) to develop an automatic classifier for tweets, which can classify tweets into our three categories of "aggression", "grief", and "other". The problem for NLP is twofold. First, the tweets are not in standard English. Second, the classification task is a difficult one. We discuss them in turn.

The tweets in our collection pose three challenges. First, they are tweets – Twitter has developed into its own genre with its unique communicative conventions, which include the use of emoticons, hashtags, URLs, non-conventional spelling, and sometimes telegraphic grammar. Second, the authors of the tweets use African American English, whose grammar differs in some respects from that of Standard American English. Finally, the language of the Chicago gangs has conventions and vocabulary, which we believe may have impacted the inter-annotator agreement. We are developing a POS tagger for this collection. We have annotated a corpus of 800 tweets with POS tags from the CMU tagset (Owuputi et al. 2013), which is adapted for tweets in general. We are using the CMU annotated corpus in conjunction with our own corpus using domain adaptation (Daumé 2007). We have also glossed some tweets into Standard American English and used the GIZA++ alignment tool to automatically derive an alignment between words and phrases in our tweet corpus and their Standard equivalent. We use this alignment to derive a glossary.

We then use these basic NLP tools to build a classifier which can detect tweets related to aggression and grieving. In addition to lexical features, POS, and features derived from emoticons and hashtags, we use information about the emotions expressed in the tweets. To obtain this information, we use the Dictionary of Affect in Language (Whissell, 1989), which has been used extensively in sentiment analysis. We use the glossary obtained through the automatic alignment, which proved to outperform other sources such as UrbanDictionary or Wiktionary. We then use an SVM to train a classifier on our data.

The NLP is work in progress, and we have summed up our ongoing work to show how this new data set can be used. We will present the details and a full evaluation in future publications.

### 4. CONCLUSION

We were motivated by increasing rates of gang violence in American cities to develop a collaboration between social work researchers and data scientists aimed at understanding the connection between these violent acts and language and communication on social media Web sites. However, we found few resources or corpora based on the unique language and communication style of the urban gang-affiliated youth of color who are frequently part of these situations.  In this paper, we describe the interaction between an in-depth textual analysis approach, focused on developing a social theory, and a data-driven machine-learning approach to developing natural language processing systems to detect that a social media post could lead to real-world violence.

In this project, we used the products of the in-depth textual analysis, a set of Twitter posts and a coding system, as an annotation for a natural language corpus intended for building systems to process these kinds of posts.  Rounds of qualitative coding were refined with feedback by data scientists using the codes, resulting in a specialized dataset relevant to automating the analysis of social media posts to detect aggressive language and potentially violent situations.  The interaction allowed the data scientists to develop tools in a methodologically systematic ways but also allowed the social work researchers to develop an analysis process that produced data that the data scientists could better use to build automated tools, overcoming challenges related to the small number of data items produced by an in-depth, textual analysis.  In future work, we will develop computer-based social media analysis tools based on our NLP systems.  We intend to provide our tools to government and community organizations focused on reducing gang violence, and we will continue to use DUVAA with larger and more diverse datasets for building social media analysis systems that can be used to curb gang violence.

5.  ACKNOWLEDGEMENTS

We would like to acknowledge Eddie Bocanegra, Ishmeel Harris, Robert Kwiatkowski, Terra Blevins and Maha Ahmed for their support of the analysis and input on this study.

6.  REFERENCES

Conway, M. ( 2006). The subjective precision of computers: A methodological comparison with human coding in content analysis. *Journalism and Mass Communication Quarterly*, 83(1), 186-200.

Daumé, Hal, III (2007). Frustratingly easy domain adaptation. In *Proceedings of the 45th annual meeting of the Association for Computational Linguistics* (pp. 256–263). Association for Computational Linguistics, Prague.

Ford, H. (2014). Big Data and Small: Collaborations between ethnographers and data scientists. Big Data and Society. 1-3

Olutobi Owoputi, Brendan O'Connor, Chris Dyer, Kevin Gimpel, Nathan Schneider and Noah A. Smith (2013).  Improved Part-of-Speech Tagging for Online Conversational Text with Word Clusters. In *Proceedings of NAACL 2013*.

Nacos, B.L., Shapiro, R.Y., Young, J.T. Fan, D. P., Kjellstrand, T., &McCaa, C. ( 1991). Content analysis of news reports: Comparing human coding and computer-assisted method. *Communication*, 12, 111-128.

Patton, D., Eschmann, R., and Butler, D. Internet banging: New trends in social media, gang violence, masculinity, and hip hop. *Computers in Human Behavior 29*, 5 (2013), 54–59

Patton, D. U., Sanchez, N., Fitch, D., Macbeth, J.,& Leonard, P. (2015). I Know God's Got a Day 4 Me Violence, Trauma, and Coping Among Gang-Involved Twitter Users. *Social Science Computer Review*, 0894439315613319.

Patton, D. U., Leonard, P., Cahill, L., Macbeth, J., Crosby, S., & Brunton, D. W. (2016a). "Police took my homie I dedicate my life 2 his revenge": Twitter tensions between gang-involved youth and police in Chicago. *Journal of Human Behavior in the Social Environment*, 1-15.

Patton, D. Lane, J., Leonard, P., Macbeth, J., Smith-Lee, J. ( 2016b). Gang Violence on the digital street: Case study of a South Side Chicago gang members Twitter communication. 1-19.

Whissell, C. ( 1989). The Dictionary of Affect in Language," Emotion: Theory, Research and Experience. The Measure of Emotions, R. Plutchick and H. Kellerman, eds., vol. 4, pp. 113-131, Academic Press.